\documentclass[12pt]{article}
\usepackage{graphicx}
\usepackage{subcaption}

\usepackage{hyperref}

\usepackage{gensymb}

\newcommand\pubnumber{}
\newcommand\pubdate{\today}

\def\napoli{School of Physics and Astronomy\\
Queen Mary University of London, E1 4NS London, UK}

\def\Title#1{\begin{center} {\Large #1 } \end{center}}
\def\Author#1{\begin{center}{ \sc #1} \end{center}}
\def\Address#1{\begin{center}{ \it #1} \end{center}}

\newcommand\pubblock{\rightline{\begin{tabular}{l} \pubnumber\\
         \pubdate  \end{tabular}}}
\newenvironment{Abstract}{\begin{quotation}  }{\end{quotation}}
\newenvironment{Presented}{\begin{quotation} \begin{center} 
             PRESENTED AT\end{center}\bigskip 
      \begin{center}\begin{large}}{\end{large}\end{center} \end{quotation}}
\def\Acknowledgements{\bigskip  \bigskip \begin{center} \begin{large}
             \bf ACKNOWLEDGEMENTS \end{large}\end{center}}

\def\beq{\begin{equation}}
\def\eeq#1{\label{#1}\end{equation}}
\def\eeqn{\end{equation}}

\def\beqa{\begin{eqnarray}}
\def\eeqa#1{\label{#1}\end{eqnarray}}
\def\eeqan{\end{eqnarray}}

\let\bar=\overbar

\def\Dslash{\not{\hbox{\kern-4pt $D$}}}
\def\dslash{\not{\hbox{\kern-2pt $\del$}}}

\def\msb{{\bar{\ssstyle M \kern -1pt S}}}

\begin{document}
\begin{titlepage}
\pubblock

\vfill
\Title{SNO+ scintillator cocktail studies using an ${}^{90}$Y source}
\vfill
\Author{Evelina Arushanova\footnote{e.arushanova@qmul.ac.uk}, Jeanne R. Wilson\footnote{j.r.wilson@qmul.ac.uk}}
\Address{\napoli}
\vfill
\begin{Abstract}
We present the design of an ${}^{90}$Y calibration source and its manufacturing procedure, that has been implemented in the University of Sussex radioactive laboratory. The radioactive source was first tested at the University of Sussex using a small scintillator cocktail sample. Further measurements were performed at the University of Pennsylvania using a larger volume of the scintillator cocktail. The results of both studies are presented and discussed.
\end{Abstract}
\vfill
\begin{Presented}
NuPhys2015, Prospects in Neutrino Physics\\
Barbican Centre, London, UK, December 16 -- 18, 2015
\end{Presented}
\vfill
\end{titlepage}
\def\thefootnote{\fnsymbol{footnote}}
\setcounter{footnote}{0}

\section{Introduction}

The SNO+ experiment probes a number of rare physics processes with the main focus on searching for neutrinoless double beta decay of ${}^{130}$Te during the tellurium loaded  scintillator phase. The detector is located in SNOLAB, located in Creighton Mine in Sudbury, Canada with approximately 2\,{\rm km} of rock overburden. The SNO+ detector is an acrylic vessel, 6m in radius, containing 780\,{\rm tonnes} LAB PPO liquid scintillator. Scintillator light from low energy interactions is observed by  about 9400 PMTs surrounding the acrylic vessel. 

In order to produce reliable results, a broad calibration system has been developed, including an optical calibration system and deployed radioactive sources, specifically ${}^{60}\mathrm{Co}$, ${}^{57}\mathrm{Co}$, ${}^{48}\mathrm{Sc}$, ${}^{24}\mathrm{Na}$, ${}^{16}\mathrm{N}$ and AmBe, which are gamma-emitters, covering an energy scale from 0.1\,{\rm MeV} to 6\,{\rm MeV} \cite{Andringa:2015tza}.  AmBe is also a source of neutrons.

It was proposed in \cite{JRWilson2011} to add a pure beta-emitter ${}^{90}$Y-isotope as a calibration source, which will verify the simulated detector model and energy reconstruction algorithms of electron-like events. The advantage of ${}^{90}$Y is the relatively short half-life time of 64\,{\rm hours} which reduces the risk of long-term contamination of the detector. The high end-point energy of 2.24\,{\rm MeV} of the decay allows to study the energy region close to the neutrinoless double beta decay.  
 
\section{ ${}^{90}$Y calibration source development} 

In order to use a calibration source, its geometry has to be understandable and easily modelled. The goal is to create a point source of beta radiation which can be achieved using a small cylindrical shaped container and a droplet of ${}^{90}$Y inside it. We found the best option was to use a micro capillary, available from various suppliers in a range of diameters and materials. A Monte Carlo study, using SNO+ RAT software was performed to define an appropriate geometry. The study shows that betas from the decay of ${}^{90}$Y are slightly less attenuated by glass than by quartz. An additional advantage of using glass is that its melting point is much lower than of quartz. We simulated different capillary diameters, from 0.5\,{\rm mm} to 2\,{\rm mm} and found that the ${}^{90}$Y contained within an outer diameter of 1.2\,{\rm mm} and inner diameter of 1\,{\rm mm} still behaved as a point source and suffered minimal attenuation in the glass. At the same time such a diameter is preferable from practical approach. The amount of  ${}^{90}$Y, and therefore the height of the droplet, can not be large, as it stops being approximated by a point, and secondly attenuates the  electrons more. After optimization we  chose to inject 2\,{\rm$\mu$L} of ${}^{90}$Y-source. 

After defining the parameters of the calibration source, we designed the manufacturing procedure, taking into account safety of personnel and ease in production. Plastic elements  required for the production and secure usage of the source were designed and machined in Queen Mary University of London. The production of the calibration source took place in the University of Sussex radioactive laboratory. All work with the radioactive source was performed behind a Perspex shield inside a fume cabinet.

\begin{figure} 
\begin{subfigure}{0.5\textwidth}
\includegraphics[scale=0.087]{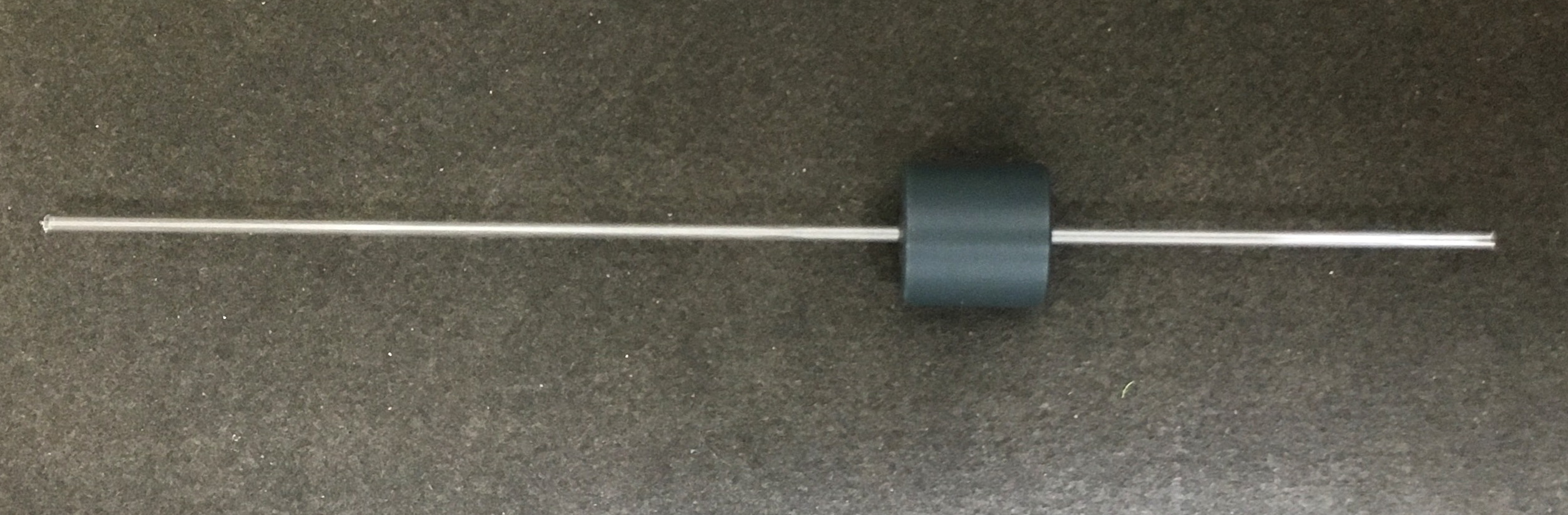}
\caption[]{The capillary with the glued holder } \label{capillary}
\end{subfigure}
\hspace*{\fill} 
\begin{subfigure}{0.5\textwidth}
\includegraphics[scale=0.1]{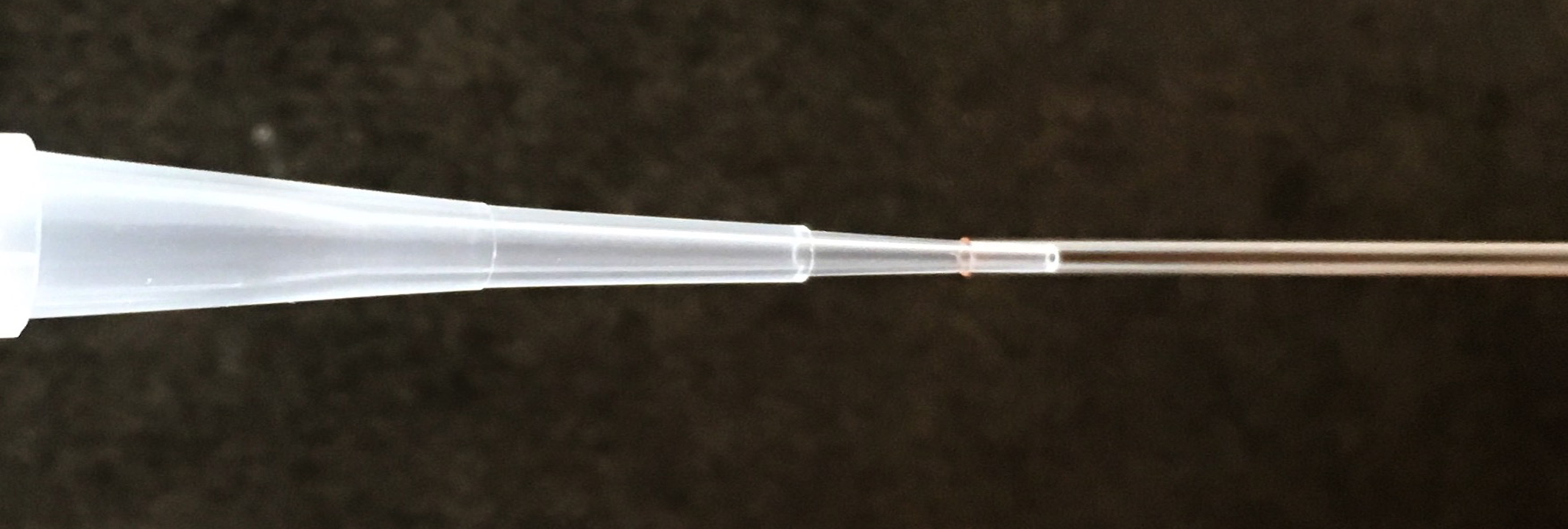}
\caption[]{The capillary and the end of the pipette} \label{tip}
\end{subfigure}
\caption{}
\end{figure} 

We used liquid ${}^{90}$Y, separated from ${}^{90}$Sr and purified at the 10$^{-7}$ level by PerkinElmer Inc. for medical treatments. Prior to injection of ${}^{90}$Y a plastic holder was glued to the capillary, Figure~\ref{capillary}. In order to safely insert the ${}^{90}$Y droplet, the capillary was placed into a plastic stand supported by the holder. A Gilson pipette was used to injected 2\,{\rm$\mu$L} of ${}^{90}$Y source. We then wiped the capillary with a dry cotton pad and moved the droplet down with 8\,{\rm$\mu$L} of air, injected from the pipette with a new end. The accuracy in this procedure is very important to avoid accidental breakage, as the diameter of a plastic end of the pipette is only slightly smaller than the inner diameter of the capillary, Figure~\ref{tip}. Inaccurate air pipetting can split the ${}^{90}$Y  into small droplets and hence the source can't be considered  as a point source. After the droplet is placed at a satisfactory position, the edges of the capillary are sealed using a butane gas torch. The seals are visually inspected under magnification; a good seal is accurate and has no glass bubbles. It was next soaked in a water bath, that was then tested for contamination, using a Geiger counter. Next the clean capillary was placed into a vacuum canister air pump storage box. If the droplet either moves or splits into smaller droplets under vacuum, this indicates a poor seal and must be repeated

We prepared two calibration sources to perform the tests in the University of Sussex and in the University of Pennsylvania. The capillary with the source was shipped in a custom designed thick acrylic container, that protected it from damage.

\section{Measurements at the University of Sussex}

Initial tests were performed in the radiation laboratory at the University of Sussex. The experimental setup consists of two Hamamatsu 1'' PMTs H10721-210 PMTs and a disk shaped vessel with diameter 6\,{\rm cm} filled with LAB PPO inside a dark box, and a Tektronix MSO2024 oscilloscope. 

\begin{figure} 
\centering
\includegraphics[scale=0.1]{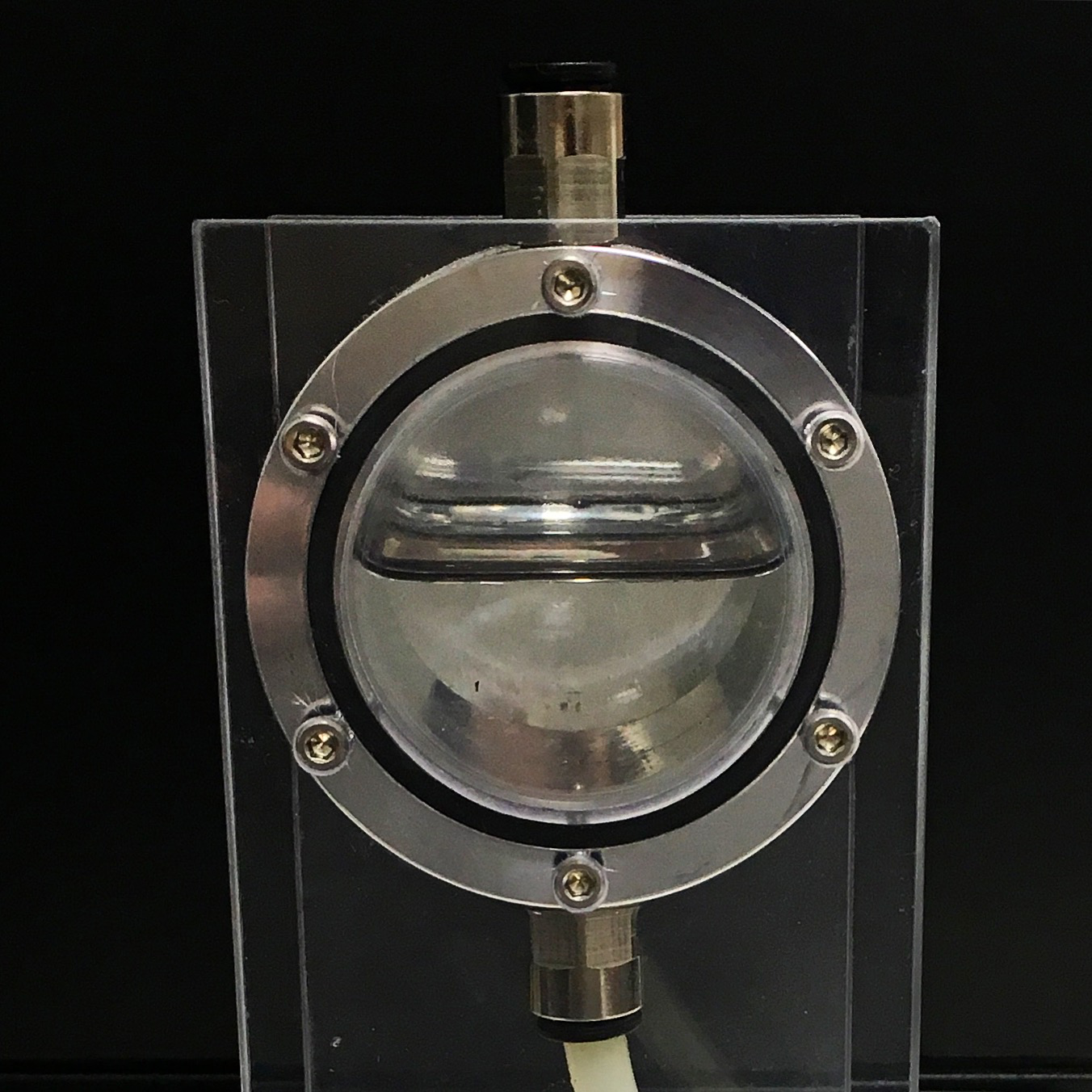}
\caption{Disk-shaped vessel filled with LAB PPO. }
\label{sussex_vessel}
\end{figure} 

The capillary is placed inside the scintillator filled vessel, Figure~\ref{sussex_vessel}. Due to the small volume of the vessel and high decay rate of the ${}^{90}$Y the contribution from cosmic ray muons is negligible. Signals were readout via the oscilloscope and analysed by custom written programs using LABVIEW and Python. 

\begin{figure} 
\begin{subfigure}{0.5\textwidth}
\includegraphics[scale=0.35]{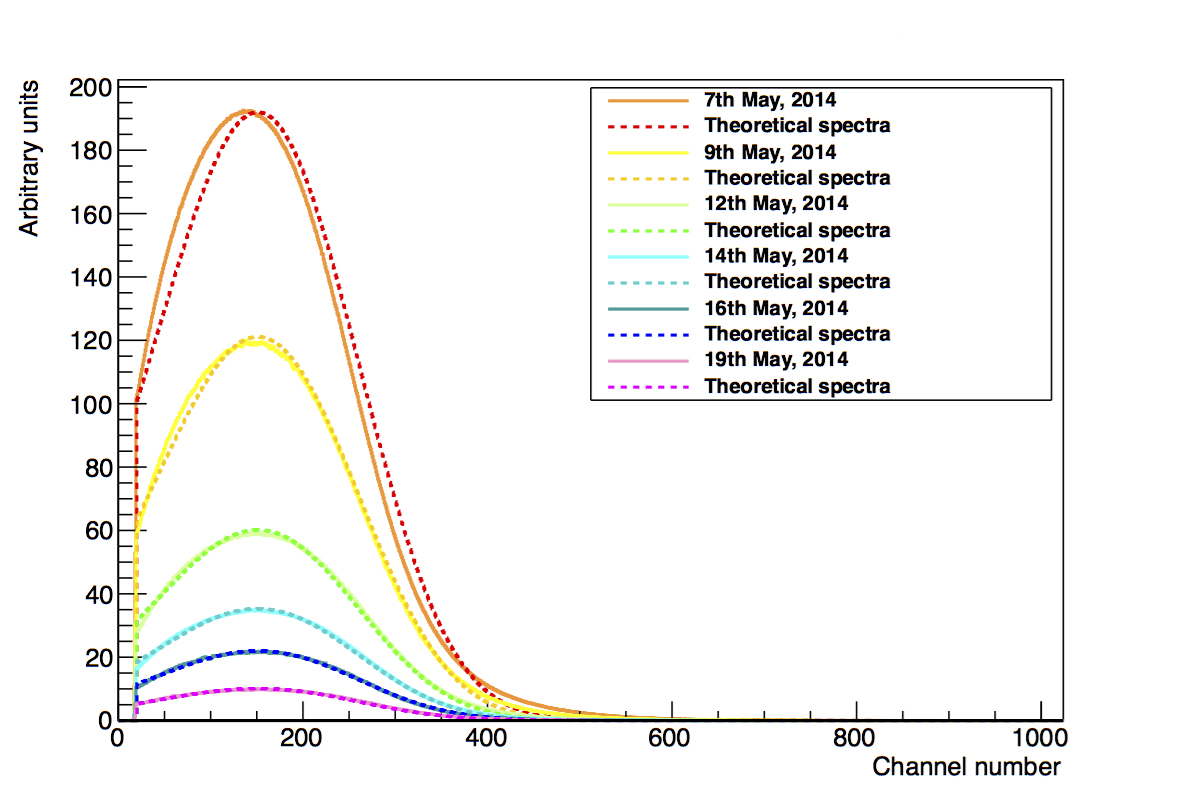}
\caption[]{Spectra of ${}^{90}$Y decay over time} \label{rainbow}
\end{subfigure}
\hspace*{\fill} 
\begin{subfigure}{0.5\textwidth}
\includegraphics[scale=0.35]{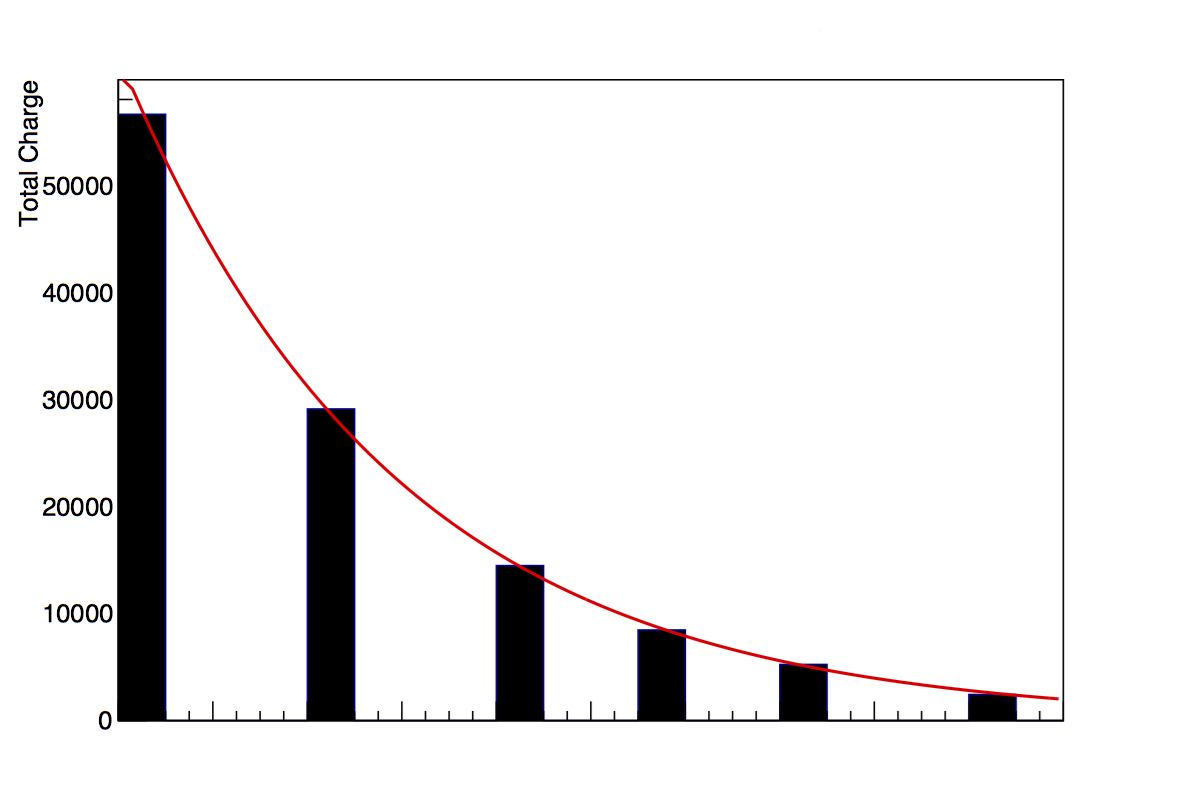}
\caption[]{Count rates of ${}^{90}$Y decay over time} \label{decay}
\end{subfigure}
\caption{}
\end{figure} 

Data from several days agrees with the theoretically predicted spectrum of ${}^{90}$Y decay, Figure~\ref{rainbow}. During the first day of data taking, when the activity of the source was high, we observed pile up between multiple decays in the same readout window. The count rate over time also agrees with the expected half-life, Figure~\ref{decay}, confirming that we can observe ${}^{90}$Y decay betas from the source with minimal attenuation.

\section{Measurements at the University of Pennsylvania}

Using a large spherical acrylic vessel (diameter 40\,{\rm cm}) at the University of Pennsylvania we were able to make more precise measurements to study the scintillator cocktail properties. The source capillary was placed in the tank using a custom designed mount. Further data was collected with a ${}^{60}$Co test disk source attached to the spherical vessel from the outside to stop betas from entering the scintillator volume. The experimental setup includes five Hamamatsu PMTs: a ETL-9354KB PMT that was used as the trigger PMT,  two R11780-HQE PMTs and two R1408 PMTs, that have been used in the SNO experiment. To accurately simulate the PMT response, the single photoelectron distribution was measured and then convolved with the simulated charge distribution from ${}^{90}$Y decay. The final modelled charge distribution was  fit to the obtained data. Using the echidna software designed by the University of Sussex and Queen Mary University of London for fitting and limit setting tasks. The parameters of the fit, including a charge scale and an offset, have been applied to simulations of ${}^{60}$Co and compared to obtained data. Poor convergence in the fit indicated discrepancies in the scintillator model, triggering further study. 

\section{Conclusion and future directions}

These studies have contributed to improve modelling of the scintillator response and validated this ex-situ source calibration technique. However these measurements were challenging, and due to the delicate nature of the capillary source, we conclude deployment of such a source within the SNO+ detector represents too high risk to detector contamination, despite the short source half-life. One possible approach is to contain the capillary within a secondary scintillator filled acrylic sphere for deployment into SNO+, but this study will be analysing this particular scintillator, rather than the scintillator volume of the detector.

\Acknowledgements
This research was supported by ERC grant 278310 under the FP7 framework. We are grateful to the University of Sussex group for providing access to their radiation laboratory and the University of Pennsylvania group for helping to perform ${}^{90}$Y source measurements in various scintillator cocktails.

\end{document}